\documentstyle[12pt]{article}
\begin{document}

\begin{titlepage}
\begin{flushright}
       {\bf UK/95-11}  \\
 Sept. 1995   \\
 hep-lat/9510046 \\
\end{flushright}
\begin{center}

{\bf {\LARGE Comments on Lattice Calculations of \\
\bigskip
Proton Spin Components }}

\vspace{1cm}

{\bf Keh-Fei Liu
 \footnote{ Report for the workshop on ``Future Physics at HERA'',
 Sept. 25-26, 1995}}\\ [0.5em]
 {\it  Dept. of Physics and Astronomy  \\
  Univ. of Kentucky, Lexington, KY 40506}

\end{center}

\vspace{0.4cm}

\begin{abstract}

Comments on the recent lattice QCD calculations of the
flavor-singlet axial coupling constant
$g_A^0$ and individual quark and gluon spin contributions to
the proton spin is given. I point out the physics learned from these
calculations as well as some of the lessons and pitfalls.

\bigskip

\end{abstract}

\vfill

\end{titlepage}

Recent experiments on polarized deep inelastic lepton-nucleon
scattering from SMC~\cite{smc94} and E143~\cite{e14395} have
confirmed the finding of the earlier EMC~\cite{emc88} results that
the flavor-singlet $g_A^0$ is small~\cite{bek88}.
In the deep inelastic limit, the integral of the polarized
structure function is related to the forward matrix elements of
axial currents from the operator product expansion~\cite{kod80}.
Combined with the neutron and hyperon decays, the
flavor-singlet axial coupling $g_A^0$ is extracted. Since the
axial current is the canonical spin operator, $g_A^0$ is thus
the quark spin content of the nucleon; i.e.
$g_A^0 = \Delta u + \Delta d + \Delta s$,
where the spin content $\Delta q (q = u,d,s)$ is
defined in the forward matrix element of the axial current,
 $\langle ps|\bar{q}i\gamma_{\mu}\gamma_5 q | ps \rangle
= 2 M_N s_{\mu} \Delta q$.

The fact that $g_A^0$, which represents the quark
spin contribution to the proton spin, is found to be much smaller
than the expected value of unity from the
non-relativistic quark model or 0.75 from the SU(6) relation (
3/5 of the isovector coupling $g_A^3 = 1.2574$)
has surprised physicists who have taken the valence quark
model for granted.
In view of the exciting puzzle presented by the experiment,
the gross failure of the valence quark model, and its possible
manifestation of the axial anomaly besides $\pi^0 \rightarrow \gamma
\gamma$, several lattice QCD calculations
were carried out in the last 6 years as an attempt to meet the challenge
of explicating the phenomenon directly from QCD.

   In the following, I shall summarize the progress that has been
made in this direction and point out the pitfalls and/or shortcomings
in each of the calculations reviewed here. There are three approaches
to address the issue of the proton spin components. The first is the
calculation of the gluon spin content, the second is the calculation of
the axial anomaly, and the third is the calculation of the
quark spin content from the axial-vector current.

\section{Gluon Spin Content}

\noindent
Ref \cite{man90}: Jeffrey E. Mandula, ``Lattice Simulation of the
Anomalous
Gluon Contribution to the Proton Spin'', Phys. Rev. Lett. {\bf 65},
1403 (1990).  \\

This is the first attempt \cite{man90} to calculate the gluon spin
content as a way of understanding the EMC experiment.

\subsection{Calculational details}

\begin{itemize}
\item Lattice parameters: 204 $6^3 \times$ 10 lattice at $\beta
      = 5.7$ and the Wilson mass parameter $\kappa = 0.162$.
\item This a quenched approximation (without dynamical fermion loops)
      with lattice spacing $a \sim 1.0\, GeV^{-1}$.
\item The gluon spin operator is the lattice version of the
      anomalous current.

 \begin{equation}  \label{ac}
 K_{\mu} =  \varepsilon_{\mu\nu\lambda\rho} Tr A^{\nu}(G^{\lambda
    \rho} - 2/3 A^{\lambda} A^{\rho})
 \end{equation}

 which is a particular case of the Chern-Simons form \cite{cs74}.
 The gluon spin fraction is defined as
 \begin{equation}
 \langle ps|K_{\mu}|ps\rangle = 2 M_N  s_{\mu} \Delta g.
 \end{equation}
\item Since the current is not gauge invariant, the calculation
  is done in the $A_4 = 0$ gauge where the second term in $K_i$
  (i = 1,2,3) vanishes.
 \item  It turns out that the signal is very noisy and only an
  upper bound is given, i.e. $\Delta g \leq |0.5|$. Since the
  original motivation of the work is to evaluate the possible gluon
  spin mixture in the quark axial-current matrix element via the triangle
  diagram \cite{et88,ar88,ccm88}, the author gives the upper limit
  of the mixture $ (3\alpha_s/2\pi)|\Delta g| \leq 0.05$.

\end{itemize}

\subsection{Comments}

 \begin{itemize}
 \item
  This work has drawn several comments : \\
  Efremov, Soffer, and  T\"{o}rnqvist \cite{est91}
  assert that what is calculated in \cite{man90} contains a contribution
  from the ``ghost'' pole. On the other hand, Manohar \cite{mano91}
  points out that even in the forward direction, $\Delta g$ may still
  depend on the gauge direction $\eta$. Mandula replies \cite{man91}
  that ``while gauge-variant singularities and $\eta$ dependence are
  kinematically allowed in the forward matrix element
  $\langle ps|K_{\mu}|ps\rangle$, they are not expected to occur,
  on either dynamical, symmetry, or invariance grounds.''
  \item
  We believe that the issue of ``ghost'' pole is a red herring. The
  ``ghost'' was invented by Veneziano \cite{ven79} for the convenience
  of taking care of a minus sign in the topological susceptibility
  which may well come from a contact term. Unlike the claim of Efremov,
  Soffer, and T\"{o}rnqvist that ``This ghost pole is necessary for
  the resolution of the U(1) problem in QCD'', Witten's resolution of
  the U(1) problem in terms of the topological susceptibility does
  not depend on the ``ghost''. Furthermore, it has been shown
  \cite{liu92} that one can adapt Veneziano's approach without
  invoking the ``ghost''. On the other hand,
  Manohar does have a point, especially in view of the fact that
  it has been shown by Cronstr\"{o}m and Mickelsson \cite{cm83} that
  under gauge transformations,

  \begin{equation}
  K_{\mu} \longrightarrow K_{\mu} + 2 \varepsilon_{\mu\nu\alpha\beta}
  \partial^{\nu}Tr A^{\alpha}U^{-1}\partial^{\beta}U
  + \partial^{\nu}H_{\mu\nu}(U).
  \end{equation}

  The last term is not sensitive to
  the local deformation of U and can be considered a topological
  current. For large gauge transformation where the topological
  (or instanton) number changes, the forward matrix element
  \begin{equation}
  \lim_{q \rightarrow 0} \langle p+q|\partial^{\nu}H_{\mu\nu}(U)|p\rangle
 = \lim_{q \rightarrow 0} iq^{\nu}\langle p+q|H_{\mu\nu}(U)|p\rangle
   \end{equation}
   does not vanish \cite{jm90,kp92}. Thus the calculation of the
   gluon spin content using the anomalous current in eq. (\ref{ac})
   is gauge-dependent even for forward matrix element.

\item
   It turns out that in the temporal axial gauge, $A_4 = 0$, the
anomalous current $K_i$ in eq. (\ref{ac}) coincides with the gluon spin
operator $2 Tr (A \times E)^i$ \cite{jm90}. Hence, modulo
gauge-dependence, what is calculated in \cite{man90} is the gluon spin
content.

\end{itemize}

\section{Flavor-Singlet $g_A^0$ from Axial Anomaly}

  In this approach, one calculates the following matrix element
\begin{equation}  \label{A}
A = \lim_{q \rightarrow 0} \frac{i|s|}{q.s}
\langle p,s|N_f \frac{\alpha_s}{2\pi} G_{\mu\nu}^a
\tilde{G}^{a\mu\nu}|p',s\rangle.
\end{equation}

  The basis for calculating A instead of the matrix element of
the axial current directly can be seen as follows.
  Inserting the anomalous Ward identity

 \begin{equation}\label{awi}
\partial^{\mu} A_{\mu}^0 = 2 \sum_{f=1}^{N_f} m_f \overline{q}_f
i\gamma_5 q_f + N_f \frac{\alpha_s}{2\pi} G_{\mu\nu}^a
\tilde{G}^{a\mu\nu},
\end{equation}

 between proton states, one obtains

 \begin{equation}\label{ga}
\overline{u}(p) i\gamma_5 u(p') [2M g_A^0 (q^2) + q^2 h_A^0 (q^2)]
=<p|2mP + 2N_f\/q|p'>,
\end{equation}
where $P$ is the pseudoscalar current, i.e. $mP =
 \sum_{f=1}^{N_f} m_f \overline{q}_f
i\gamma_5 q_f $ and {\it q} is the topological charge operator,
$q = (\alpha_s/4\pi) G_{\mu\nu}^a \tilde{G}^{a\mu\nu}$.

Since there is no Goldstone pole at the chiral limit in the
U(1) channel, one can take the $q^2 \rightarrow 0$ and the
$m \rightarrow 0$ limits and drops the $h_A$ and $P$ terms in
eq. (\ref{ga}) to obtain

\begin{equation}  \label{gaA}
g_A^0 = \Delta\Sigma = A
\end{equation}

Thus one can calculate the matrix element
of the axial anomaly to obtain $\Delta\Sigma$. One will
not be able to calculate $\Delta u$, $\Delta d$, and $\Delta s$
separately in this approach though.

\subsection{Quenched Approximation}

\subsubsection {Two calculations}

\begin{enumerate}
\item Ref \cite{acd94}: B. All\'{e}s, M. Campostrini, L. Del Debbio,
A. Di Giacomo,
H. Panagoulos, and E. Vicari, ``The Proton Matrix Element of the
Topological Charge in Quenched QCD'', Phys. Lett. {\bf B336},
248 (1994).

\begin{itemize}
\item Lattice parameters: 100 $ 16^3 \times 32$ lattice at
$\beta = 6.0$. Smeared source for the nucleon in the Coulomb gauge.
Wilson hopping parameters $\kappa = 0.153, 0.154, 0.155$ are used for
the quark propagators.
\item Nonperturbative method is used to determine the finite lattice
renormalization constant.
\item The signal is zero within errors. An upper bound of
 the matrix element A in eq. (\ref{A}) is set to be 0.4.
\end{itemize}

\item
Ref \cite{gm94}: R. Gupta and J.E. Mandula, ``Matrix Elements of the
Singlet Axial
Vector Current in the Proton'', Phys. Rev. {\bf D50}, 6931 (1994).

\begin{itemize}
\item
Lattice parameters: 35 $16^3 \times 40$ lattice at $\beta = 6.0$.
The quark masses used are $\kappa = 0.154$ and 0.155, corresponding
to pions of mass 700 and 560 MeV, respectively. The Wuppertal source
is used for the nucleon, hence it is a gauge invariant calculation.
\item
Again, the authors found that the data are statistically
indistinguishable from zero.
\end{itemize}

\subsubsection{Comments}

\begin{itemize}
\item
Irrespective of the fact that no signal is seen in these calculations,
it is pointed out emphatically by Gupta and Mandula \cite{gm94} that
this approach which relies on the use of the axial anomaly fails in the
quenched approximation. This can be see as follows.
What is derived in eq. (\ref{gaA}) depends on the fact that there
is no massless Goldstone mode in the U(1) channel at the chiral limit.
However, this is not realized in the QUENCHED approximation where
the would be Goldstone bosons are present. It is clearly seen in the
study of the anomalous Ward identity for $g_A^0$ \cite{liu92} that
the induced pseudoscalar form factor $h_A$ has a pole structure
$\frac{1}{q^2 - m_{\eta_0}^2}$ where $m_{\eta_0}$ is the would-be
Goldstone meson mass, while the pseudoscalar current 2mP
has a pole structure $\frac{m_{\eta_0}^2}{q^2 - m_{\eta_0}^2}$.
Combined, they give the structure $\frac{q^2 - m_{\eta_0}^2}
{q^2 - m_{\eta_0}^2}$ which tends to unity at the $q^2 \rightarrow 0$ and
$m \rightarrow 0$ limits. In other words, in the quenched approximation,
$g_A^0$ is not simply related to A as in eq. (\ref{gaA}).
The pseudoscalar current 2mP also contributes in this case even in the
chiral limit as is in the Goldberger-Treiman relation.

\item
Gupta and Mandula \cite{gm94} further suggest that due to the presence
of the double/single would-be Goldstone pole in the pseudoscalar current
/axial anomaly, the result from the quenched approximation diverges
at the chiral limit and thus one will not be able to
utilize the anomalous Ward identity to calculate $g_A^0$. Fortunately,
it can be seen from our earlier study \cite{liu92} that the double pole
in the pseudoscalar current cancels the single pole from the axial
anomaly. This is due to the fact that the topological susceptibility
vanishes at the chiral limit so that the vacuum has no $\theta$
dependence. Furthermore,  the remaining single would-be Goldstone pole
(at $q^2 = 0$) will be cancelled by the quark mass in the pseudoscalar
current 2mP. Thus every term in the expression of eq. (\ref{ga}) is
finite in the chiral limit of the quenched approximation. We conclude
that one can still utilize the anomalous Ward identity in the quenched
approximation, provided the pseudoscalar current is evaluated in addition
to the axial anomaly term.
\end{itemize}
\end{enumerate}

\subsection{Calculation with Dynamical Fermions}

\noindent
Ref. \cite{agh94}: R. Altmeyer, M. G\"{o}ckler, R. Horsley, E. Laermann,
and G. Schierholz, ``Axial Baryonic Charge and the Spin Content of
the Nucleon: a Lattice Investigation'', Phys. Rev. {\bf D49}, R3087
(1994).

\subsubsection{Calculational Details}

\begin{itemize}
\item
Lattice parameters: $16^3 \times 24$ dynamical fermion lattice at
$\beta = 5.35$ with four flavors of staggered fermions with mass
$ma = 0.01$. 85 gauge configurations separated by 5 trajectories are
used. The lattice spacing determined by the $\rho$ mass is 0.14 fm.
\item
The result of eq. (\ref{A}) calculated at $|\vec{q}| \sim
500$ MeV gives $A = 0.18 \pm 0.02$ which the authors identify with
$\Delta \Sigma$ through eq. (\ref{gaA}).
\end{itemize}

\subsubsection{Comments}
\begin{itemize}
\item
 This is the first calculation which uses the dynamical fermion gauge
 configurations and actually sees a signal which is quite commendable.
 Moreover, the result agrees with the experimental finding.
 \item
 However, this calculation is done at finite q and finite quark mass m.
 As we alluded to earlier, the induced pseudoscalar
 form factor $h_A$ will contribute at finite q (NB. eq. (\ref{ga})) and
 the pseudoscalar current 2mP will contribute at finite m. As long as
 the quark mass (in both the valence and the fermion loops) and the
 momentum transfer q are not extrapolated to 0, the relation in
 eq. (\ref{gaA}) will always be subjected to these systematic errors.
 In this approach, one can not separately calculate $\Delta u$,
 $\Delta d$, and $\Delta s$. Furthermore, just knowing $g_A^0$
 through the matrix element of the anomaly does not teach us why
 $g_A^0$ is as small as it.
\end{itemize}

\section{Quark Spin Content from the Axial-Vector Current}

   This is the direct way of calculating the quark spin content, since
it is defined by the forward axial current matrix
element and one can calculate it flavor by flavor. The only problem is
that in addition to the connected insertion, there is a disconnected
insertion. The disconnected insertion requires the calculation of
quark loops which is prohibitively time-consuming if one is to
invert the whole quark matrix directly. This is the reason why
earlier attempts were made to calculate the axial anomaly in order to
circumvent this numerical difficulty. It turns out that there have been
three calculations which tackle this problem with different numerical
techniques.

\subsection{Quark Loops in a Smaller Box}

\noindent
Ref. \cite{mo93}: Jefferey E. Mandula and Michael C. Ogilvie, ``A
New Technique for Measuring the Strangeness Content of the Proton on
the lattice'', Phys. Lett. {\bf B312}, 327 (1993).

\subsubsection{Technical Details}
\begin{itemize}
\item
Lattice parameters: 16 $16^3 \times 24$ quenched lattice at
$\beta = 5.7$. The valence quark masses are about 0.6, 0.2, and 0.1 GeV
corresponding to Wilson $\kappa = 0.140, 0.160$, and 0.164. The
strange mass in the loop is about 0.16 GeV ($\kappa = 0.162$).
\item
The quark propagator for the loop is calculated in a $9^4$ box and the
loop is evaluated on a time slice for every other site in each spatial
direction. This saves a factor $\sim$ 500 in computer time comparing
to the straight forward calculation and thus makes the project feasible.
\item
$\Delta s$ tends to be negative for light valence quarks in the nucleon.
However, no plateau is reached to give a final result.
\end{itemize}

\subsubsection{Comments}
This technique should work. The devil is in the statistics.

\subsection{Volume Source Technique}

\noindent
Ref. \cite{fko95}: M. Fukugita, Y. Kuramashi, M. Okawa, and A. Ukawa,
``Proton Spin Structure from Lattice QCD'', Phys. Rev. Lett. {\bf 75},
2092 (1995).

\subsubsection{Technical Details and Results}
\begin{itemize}
\item
Lattice parameters: 260 $16^3 \times 20$ quenched lattice at $\beta
= 5.7$. The Wilson hopping parameters $\kappa = 0.160, 0.164$,
and 0.1665 are used. The strange quark corresponds to $\kappa = 0.160$.
\item
The quark loops are calculated with the volume source technique
\cite{kfm93} which
relies on the cancellation of gauge non-invariant part through gauge
averaging. The time slice from which the nucleon source emerges
is fixed in Coulomb gauge.
\item
The results are summarized in the following table.

\begin{table}[ht]
\caption{Quark spin contents of the proton form the volume source
calculation}
\begin{center}
\begin{tabular}{ll}
 \hline
 $g_A^0 (\Delta \Sigma)$ & 0.18(10)  \\
 $g_A^3 $  & 0.985(25) \\
 $\Delta u $ & 0.638(54)  \\
 $\Delta d $ & - 0.347(46)  \\
 $\Delta s $ & - 0.109(30)   \\
 $F_A $ & 0.382(18)  \\
 $D_A $ & 0.607(14)  \\
 $F_A/D_A$ & 0.629(33)  \\
 \hline
 \end{tabular}
\end{center}
 \end{table}
\end{itemize}

\subsubsection{Comments}
\begin{itemize}
\item
The result of $\Delta \Sigma$ agrees with experiments well and its
smallness is understood as due the sea-quarks which give negative
contributions to $\Delta \Sigma$.
\item
$g_A^3$, $F_A$, and $D_A$ are smaller than the experiments by
$\sim$ 25 \%. Possible sources of systematic errors include the scaling
violation effect due to a fairly large lattice spacing \cite{fko95}
and the
fitting procedure with the ratio method which can give a jackknife
bias of $\sim$ 10\% as compared to the more desirable simultaneous
fitting procedure \cite{ldd94a}.
\item
Since the time slice from which the nucleon source emerges is fixed
to the Coulomb gauge, the results are gauge dependent. Furthermore,
the quark current operator $A_{\mu}$ can mix with the gauge
non-invariant anomalous current $K_{\mu}$ in eq. (\ref{ac}), a problem
\mbox{d\'{e}j\`{a} vu}. To see this, one can construct the lattice
version of $K_{\mu}$ with staples in such a way that their ends touch
the time slice which is gauge fixed or use the version given by
Mandula \cite{man90} with the sandwiching U-links lie in the gauge-fixed
time slice. These $K_{\mu}$ matrix elements do not vanish on this
lattice and become a part of the results obtained in this work.
It is then necessary to disentangle this gluon spin contribution from
the quark spin part before addressing the proton spin components.
\end{itemize}

\subsection{Stochastic Estimation with $Z_2$ Noise}

\noindent
Ref. \cite{dll95}: Shao-Jing Dong, Jean-Francois Laga\"{e}, and
Keh-Fei Liu, ``Flavor-Singlet $g_A^0$ from Lattice QCD'', Phys.
Rev. Lett. {\bf 75}, 2096 (1995).

\subsubsection{Technical Details and Results}

\begin{itemize}
\item
Lattice parameters: 24/50 $16^3 \times 24$ quenched lattice
configurations at $\beta = 6.0$ for the connected/disconnected insertion.
The Wilson hopping parameters are $\kappa = 0.148, 0.152$, and 0.154.
The strange quark corresponds to $\kappa = 0.154$.
\item
No gauge-fixing is done. Hence the calculation is gauge invariant.
The quark loops are calculated by the stochastic method with the
optimal $Z_2$ noise \cite{dl94} which is shown to give the correct
answers for the forward matrix elements of the vector and pseudoscalar
currents \cite{dll95}.
\item
Finite $ma$ correction for the quark loops of the Wilson fermion
is taken into account \cite{ll95}.

\item
The results are summarized in the following table.

\begin{table}[ht]
\caption{Axial coupling constants and quark spin contents of proton in
comparison with experiments}
\begin{tabular}{lll}
 \multicolumn{1}{c}{} &\multicolumn{1}{c}{This Work} &
 \multicolumn{1}{c} {Experiments} \\
 \hline
 $g_A^0 =\Delta u + \Delta d + \Delta s$
 & 0.25(12)& 0.22(10) \cite{smc94} / 0.27(10)\cite{e14395}  \\
 $g_A^3 =\Delta u - \Delta d$ & 1.20(10) \cite{ldd94a}
 & 1.2573(28) \\
 $g_A^8  =\Delta u + \Delta d - 2\Delta s$ & 0.61(13) &
 0.579(25) \cite{cr93}   \\
 $\Delta u $ & 0.79(11) & 0.80(6)\cite{smc94} / 0.82(6)\cite{e14395} \\
 $\Delta d $ & - 0.42(11) &  - 0.46(6)\cite{smc94} / - 0.44(6)
 \cite{e14395}  \\
 $\Delta s $ & - 0.12(1) & - 0.12(4)\cite{smc94} / - 0.10(4)
 \cite{e14395}  \\
 $F_A = (\Delta u - \Delta s)/2$ & 0.45(6) & 0.459(8)
 \cite{cr93}  \\
 $D_A = (\Delta u - 2 \Delta d + \Delta s)/2$ & 0.75(11)
  & 0.798(8) \cite{cr93}  \\
 $F_A/D_A$ & 0.60(2) & 0.575(16) \cite{cr93}  \\
 \hline
 \end{tabular}
 \end{table}
\end{itemize}

\subsubsection{Comments}
\begin{itemize}
\item
This calculation is
gauge-invariant in that no gauge-fixing is applied. Since there is
no gauge-invariant dimension-three axial operator for the gluon spin
as we discussed earlier \cite{cm83,jm90}, these results are not mixed
with the gluon spin. Only the quark spin content contributes.
\item
The physical picture of $g_A^0$ is starting to come in view.
The smallness of the quark spin content compared to
the non-relativistic value of unity is, first of all, due to
the fact that the combined relativistic effect and polarization
of the cloud-quarks reduces the connected insertion to $0.62 \pm 0.09$,
 a value very
close to $g_A^8$, i.e. $g_{A, con}^0 \simeq g_A^8$. This is because
the disconnected insertion is
almost independent of the flavors $u, d$, and $s$.  Furthermore, the
sea-quark polarization is large and in the opposite direction of
the proton spin. It is the sum of all these effects that produces
a small $g_A^0$.
\item
It is interesting to observe that the sea-quark spin contribution
is independent of the quark mass in the loop within errors. This is
reminiscent of the $\gamma_5$ current in the context of the topological
charge and susceptibility. It is suggestive of the importance of the
zero modes and the instantons.
\item
Albeit the results all agree with experiments, the systematic
errors due to the finite volume, discretization, renormalization,
and quenched approximation (which could
be as large as $7\%$ -- $20\%$\cite{ldd95}) need to be addressed
before a final precise comparison with experiments can be made.
\end{itemize}

\subsection{Moments of Polarized Structure Functions}

\noindent
Ref. \cite{ghi95}: M. G\"{o}ckler, R. Horseley, E.-M. Ilgenfritz,
H. Perlt, P. Rakow, G. Schierholz, and A. Schiller, ``Polarized and
Unpolarized Nucleon Structure Functions from Lattice QCD'',
DESY 95-128, hep-lat/9508004.

\subsubsection{Technical Details}
\begin{itemize}
\item
Lattice parameters: 400 -- 1000 $16^3 \times 32$ quenched lattice
configurations at $\beta = 6.0$. The hopping parameters are $\kappa =
0.1515, 0.153$, and 0.155.
\item
$g_A^3 = 1.07(9)$, the connected insertion part of $g_A^0$ is
0.59(5). The $g_1$ sum rule is 0.166(16) for the proton and
-0.008(09) for the neutron.
\end{itemize}

\subsubsection{Comments}
\begin{itemize}
\item
The results are from the connected insertion only. They have very
high statistics. The results on the disconnected insertions are
forth-coming.
\end{itemize}

\section{Future}

So far, the lattice calculation is only beginning to shed some
light on the quark spin content part of the proton spin. We need
to calculate the orbital angular momentum and the
gluon spin in order to complete the picture of the proton spin
composition. In the future, the gauge invariant calculation of
the axial-vector matrix elements to address the infinite volume
and continuum limits and the dynamical fermions is needed to
study the systematic errors of the quark spin content.

This work is partially supported by DOE
Grant DE-FG05-84ER40154.

\end{document}